\documentclass{ieeeaccess}
\usepackage{cite}
\usepackage{amsmath,amssymb,amsfonts}
\usepackage{algorithmic}
\usepackage{graphicx}
\usepackage{textcomp}

\usepackage{url}
\usepackage{flushend}
\usepackage{tabularx}
\usepackage{caption}
\usepackage[caption=false]{subfig}
\usepackage[colorlinks=true,linkcolor=blue,urlcolor=black,bookmarksopen=true]{hyperref}
\usepackage{bookmark}
\DeclareMathOperator*{\argmax}{arg\,max}

\newtheorem{definition}{Definition}

\def\BibTeX{{\rm B\kern-.05em{\sc i\kern-.025em b}\kern-.08em
    T\kern-.1667em\lower.7ex\hbox{E}\kern-.125emX}}

\begin{document}

\history{Date of publication xxxx 00, 0000, date of current version xxxx 00, 0000.}
\doi{10.1109/ACCESS.2017.DOI}

\title{Using Intuitionistic Fuzzy Set for Anomaly Detection of Network Traffic from Flow Interaction}
\author{\uppercase{Jinfa Wang}\authorrefmark{1},
	\uppercase{Hai Zhao\authorrefmark{1}},
	\uppercase{Jiuqiang Xu\authorrefmark{1}},
	\uppercase{Hequn Li\authorrefmark{1}},
	\uppercase{Shuai Chao\authorrefmark{1}},
	\uppercase{Chunyang Zheng\authorrefmark{1}}
}
\address[1]{School of Computer Science and Engineering, Northeastern University, Shenyang, 110819 China}

\tfootnote{This work was supported by the Fundamental Research Funds for the Central Universities under Grant 02190022117021 and N171903002.}

\markboth
{J. Wang \headeretal: Using Intuitionistic Fuzzy Set for Anomaly Detection of Network Traffic from Flow Interaction}
{J. Wang \headeretal: Using Intuitionistic Fuzzy Set for Anomaly Detection of Network Traffic from Flow Interaction}

\corresp{Corresponding author: Jinfa Wang (e-mail: jinfa.wang@mervin.me).}

\begin{abstract}

	We present a method to detect anomalies in a time series of flow interaction patterns. There are many existing methods for anomaly detection in network traffic, such as number of packets. However, there is non established method detecting anomalies in a time series of flow interaction patterns that can be represented as complex network. Firstly, based on proposed multivariate flow similarity method on temporal locality, a complex network model (MFS-TL) is constructed to describe the interactive behaviors of traffic flows. Having analyzed the relationships between MFS-TL characteristics, temporal locality window and multivariate flow similarity critical threshold, an approach for parameter determination is established. Having observed the evolution of MFS-TL characteristics, three non-deterministic correlations are defined for network states (i.e. normal or abnormal). Furthermore, intuitionistic fuzzy set (IFS) is introduced to quantify three non-deterministic correlations, and then a anomaly detection method is put forward for single characteristic sequence. To build an objective IFS, we design a Gaussian distribution-based membership function with a variable hesitation degree. To determine the mapping of IFS's clustering intervals to network states, a distinction index is developed. Then, an IFS ensemble method (IFSE-AD) is proposed to eliminate the impacts of the inconsistent about MFS-TL characteristic to network state and improve detection performance. Finally, we carried out extensive experiments on several network traffic datasets for anomaly detection, and the results demonstrate the superiority of IFSE-AD to state-of-the-art approaches, validating the effectiveness of our method.  
\end{abstract}
\begin{keywords}
	network traffic flow, flow interaction, complex network, anomaly detection, multivariate flow similarity, temporal locality, intuitionistic fuzzy set, ensemble method.
\end{keywords}

\titlepgskip=-15pt

\maketitle

\section{Introduction}
\label{sec:introduction}

Anomalies can be defined as some patterns in data that do not imitate network traffic normal behavior. Network operators frequently face a wide range of such patterns in network traffic. Most of anomalous patterns begin abnormalities due to malicious illegitimate activities under large scale traffic flow interactions, such as cyber intrusions, distributed denial of services attacks, botnet attacks, worm propagation, port malicious scanning, and brute-force attacks \cite{Bhuyan2014, Weller-Fahy2015}. Which could lead to catastrophic consequences and threaten the proper operation of networks. Anomaly detection is a method to find patterns that deviate from the expected behavior. Although network traffic profiling has become one of the most important and common means for anomaly detection in the past years\cite{Xu2007}, new and more robust detection mechanisms need to be developed as the complexity of these attacks keeps increasing as discussed in Cisco 2017 Midyear Cybersecurity report\cite{cisco2017}.

In general, the studies of network traffic profiling can be classified by their level of observation: (a) packet level, such as signature-based application detection and methods using the well known port numbers, (b) flow level statistical techniques (c) host level, such as host-profiling approaches. The above methods focus on the feature extraction, selection, and analysis on per study object (e.g. packet, flow or host), and ignore the interaction of network traffic. In this paper, the interaction refers to the dependency between two flows or relevance from one flow to another. For example the worm virus hiding in a host will firstly send a large number of scanning flows to other hosts, and then inject worm code into the vulnerable hosts depending on the scanning response flows. And in both the case of malcode and P2P, using content signature methods seem destined to fail in the face of encryption and polymorphism. Therefore, from the perspective of traffic flow interactions, we construct a complex network model to describe the interaction behaviors of large scale network traffic flows instead of traditional statistical method. Since complex network provides a powerful mechanism for capturing the interactive relationships among study objects, it has been an effective method for relational expression of structured data\cite{Liu2018,Supriya2016,wang201637}, especially the time series data. For instance, Supriya et al.\cite{Supriya2016} translated the epileptic EEG signal time series into complex network, and then used the statistical properties of complex network to detect the epilepsy. Whereas in Internet, based on complex networks theory, the complexity of Internet topologies have been widely studied \cite{Garcia-Robledo2013,wang201701}. But there is few studies about the complexity of network traffic. In 2007, a study of the social behavior of Internet hosts was presented by the Traffic Dispersion Graphs (TDGs) as a way to monitor, analyze, and visualize network traffic\cite{Iliofotou2007}. In TDGs, the edge can be defined to show different interactions between two hosts. Wang et al.\cite{Xu2014} studied social behavior similarity of Internet end-hosts based on behavioral graph analysis. In Reference\cite{Wang201807}, we studied the flow interactive behaviors on temporal locality and put forward the temporal locality complex network (TLCN) to monitor, analyze, and visualize large scale network traffic flows. But in TLCN, some network flows that are not relevant in flow content could also build connections due to TLCN definition. For this reason, by focusing on the interaction behaviors of similar flows on temporal locality, an improved complex network model (MFS-TL) for large scale network flows is developed in which the proposed multivariate flow similarity method is used to quantify the correlation among the flows on temporal locality.

Obviously, the anomaly detection for network traffic flows is to detect the anomalous MFS-TL structure. Recently years, constructing complex networks from structured data and then mining the nonconforming patterns by complex network statistical characteristics have become an effective means for anomaly detection\cite{Akoglu2015}. For instance cyber networks, fraud detection, fault detection in medical claims, engineering systems, sensor networks, climate network, and many more domains. However, one of key challenges is the inconsistent performance about multiple network characteristics to network states\cite{Rayana2016,wang201701,Ai2013}, when each of network characteristics is regarded as the constituent detector alone.

Specifically, some characteristic values present negative correlation with detection goals. In other words, this characteristic could indicate that network is normal/abnormal but in fact it is abnormal/normal. Some have non-correlation with the detection goals, i.e. the unuseful characteristics could bring non-deterministic for network states\cite{ZHU2142}. Hence it can be inferred that the good characteristics improve the detection accuracy, the bad ones strength the certainty for the opposite state, but the others bring the uncertainty for judging network state. Undoubtedly, it become an uncertain theory problem of multiple complex network characteristics to multiple states.

In this paper, the intuitionistic fuzzy set (IFS)\cite{Zhang2017} is adopted to describe above non-deterministic correlation problem. Its key is that similar networks probably share certain characteristics\cite{Supriya2016}. Different from \cite{kavitha2011}, where it is given a non-null hesitation part about the evaluation of study objects to define the indeterministic behavior, we use the hesitation degree of IFS to express the useless of unuseful characteristic, and the non-membership degree of IFS to represent the negative correlation about the characteristic to one certain network state (i.e. normal or abnormal). Meanwhile a new membership function is designed to resolve the problem of the hesitation degree being a fixed value. A distinction index is put forward for the purpose of multiple clustering intervals being mapped to two states' linguistic variables. Furthermore, the intuitionistic fuzzy weighted geometric (IFWG) operator\cite{Xu2006} is introduced to fuse multiple IFSs into one new IFS about network structure to network state to eliminate the impacts of the inconsistent performance. To obtain the detection result, we use score function and precision function to select best IFS which has maximum membership degree for a given network state (or linguistic variable). Thus based on the intuitionistic fuzzy set, the detection method of single MFS-TL characteristic (IFS-AD) and the ensemble method of multiple characteristics (IFSE-AD) are separately developed to find the abnormal MFS-TL. We apply our methods to detect the anomalies in publicly network traffic trace datasets, where it utilizes 14 network characteristic metrics. Extensive evaluation on datasets with ground truth shows that IFSE-AD outperformance the other methods. The main contributions of this paper include:
\begin{enumerate}
	\item Construct a complex network model (MFS-TL) to describe the interaction behaviors of large scale similar flows on temporal locality. The MFS-TL can monitor, analyze, and visualize network flow behaviors from the perspective of the protocol, application, flow type, and flow payload.
	\item Put forward a quantification method about MFS-TL characteristic to network state based on intuitionistic fuzzy set (IFS). In which, the proposed distinction index resolves the mapping problem from multiple clustering intervals of the IFS to two states' linguistic variables.
	\item Propose an ensemble method (IFSE-AD) for multiple characteristic IFSs to improve detection performance. The IFSE-AD resolves the inconsistent problem of MFS-TL characteristic presenting network state.
	\item Methodological identification of an appropriate multivariate flow similarity critical threshold $r_c$ and temporal locality window $\Delta w$. To evaluate the efficiency of the proposed methods in detecting the anomalies of publicly datasets including the abnormal events of the cyber intrusions, botnet attacks, distributed denial of services, brute force attack.
\end{enumerate}

The rest of this paper is organized as follows: Second \ref{sec:related-work} provides a thorough literature review of the related work. Section \ref{sec:model} constructs a complex network model (MFS-TL) to describe the interaction behaviors of large scale similar flows on temporal locality. In Section \ref{sec:method}, based on intuitionistic fuzzy set, we develope a anomaly detection method for single MFS-TL characteristic, and a ensemble detection method for multiple MFS-TL characteristics' IFSs which improves the detection performance. Furthermore, the MFS-TL parameters and statistical characteristics are analyzed, and our method performance are shown and evaluated in Section \ref{sec:results}. Finally, Section \ref{sec:conclusion} presents the concluding remarks.

\section{Related work}
\label{sec:related-work}

Network traffic anomalies are instances in data that do not conform to the behavior exhibited by normal traffic. Traffic anomalies in a network can be defined as any network events or operation that deviate from the normal network behavior. They happen due to the growing number of network-based attacks or intrusions. Nowadays, the network anomaly detection methods can be classified into the statistic-, classification-, clustering-, soft computing-, fuzzy set theory-, and combination learners-based\cite{Bhuyan2014,AHMED201619}. In general, the first step is that the raw data is reduced by one of the three important strategies, i.e. dimensionality reduction (e.g. feature selection and feature extraction), clustering, and sampling, on packet-level, flow-level or host-level. However, the selected features or quantified metrics are usually used as the individual indicator\cite{Sasaka2018,Wang201806,Wang201705,Ren2018}. 

In recent years, network traffic profiling based on the probability relationships of studied objects has been paid attention. That is because complex network has been an effective method for relational expression of structured data as complex network theory provides a powerful mechanism for capturing the interactive relationships among study objects \cite{Akoglu2015}. 
For instance the communication among the hosts. In a study of the traffic volume between each pair of hosts and traffic flows on all ports, Yin et al. \cite{Yin2004} developed the VisFlowConnect tool for visualizing network traffic flow dynamics for situational awareness that on all protocols and dynamic evolution on time. Iliofotou et al.\cite{Iliofotou2007} proposed the Traffic Dispersion Graph (TDG) to discover network-wide interactions of the hosts to monitor, analyze, and visualize network traffic. Wang et al.\cite{Xu2014} studied social behavior similarity of Internet end-hosts based on behavioral graph analysis. Guan et al. \cite{Guan2010} constructed a directed graph model from the bidirectional region flows where a node denotes one host located at either source region or destination region and the edges represent the number of packets transferred from the source host to destination host.
In addition, probability graph is used to depict the relationship of the flow attributes (as the nodes). The studies of \cite{Zhou2009,Glatz2014} presented the proposed time series graph to describe the relationships among multi-time series such as the source IP address, destination IP address, source port, destination port. Di et al. \cite{Di2013} put forward the flow graphs in the following way: each mobile phone number in records represents a user node; each server IP address in the records represents a server node; each flow record between a user node and a server node forms an edges. Tsuruta et al. \cite{Tsuruta2013} defined packets sent by coordinated malware attacks as bridge-less connected bipartite graphs. The structure-based screening is a method for extracting only packets that constitute 2-edge connected components of a bipartite graph. 
Nowadays, the visibility graph is also used in the network traffic analysis. Ye et al. \cite{Ye2014} constructed the complex network for traffic sequences based on visibility graph. Then the complex network features is used to analyze the host behavior in the traffic sequence. 

In summary, the researchers have used the complex network theory to study the interaction behaviors of packet-level and host-level, or the probability relationship of the statistical results. However, in \cite{Wang201807}, we proposed the temporal locality complex network for flow-level interaction which can effectively determine the Internet applications and discover the attack patterns. In this paper, we constructed the complex network model of multivariate flow similarity on temporal locality to improve the ability of flow interaction presenting. 

Thus, the problem of network traffic anomaly detection is converted into the anomaly detection for complex networks. In fact, the anomaly detection for complex networks has become a hot topic\cite{Akoglu2015,wang201701}. Bunke\cite{Bunke1999} used graph features generated by graph edit distance to classify the normal and abnormal. Dai et al.\cite{Dai2012} unified both positive and negative mutual dependency relationship in an unsupervised framework to detect anomalous nodes of bipartite graphs such as users-rating-products in online marketplaces, users-clicking webpages on the WWW and users referring-users in social networks. Eberle and Holder \cite{Eberle2007} discovered anomalies in graphs and patterns of varying sizes with minimal to no false positive by using the minimum description length principle and probabilistic approaches. Gunnemann et al.\cite{Gunnemann2010} proposed a method for finding homogeneous groups by joining the paradigms of subspace clustering, i.e. we determine sets of nodes that show high similarity in subsets of their dimensions and that are as well densely connected within the given graph. Li et al.\cite{Li2010} developed an iBlackhole-DC algorithm for finding black hole and volcano patterns in a large directed network. Sun et al.\cite{Sun2005} identified abnormal nodes by computing the neighborhood for each node using random walk with restarts and graph partitioning. Reference\cite{Sun2010} proposed the density-based network clustering algorithm to detect communities, hubs, and outliers in large scale undirected networks. Wang et al.\cite{Wang2011} proposed a heterogeneous review graph to capture the relationships among reviewers, reviewers and stores, and then put forward an iterative model to identify suspicious reviewers. Supriya et al.\cite{Supriya2016} made the epileptic EEG signals transform into the complex network and then used the statistical properties to detect the epilepsy.

In terms of network traffic anomaly detection, some studies based on complex network theory have been done by the researchers. For example, based on the TDG, Le et al.\cite{Le2011} had used complex network metrics, such as degree distribution, maximum degree and $dK-2$ distance, to detect anomalous network traffic. In the study of network-wide anomaly detection, Zhou\cite{ZHOU2011} detected the network anomalies based on routers' connecting relationships, i.e. he used the graph to describe the traffic feature distribution sequences and their relationships. Reference \cite{Guan2010} used the six features proposed based on the regional flow model to describe the network traffic patterns and to capture the dynamic traffic patterns, especially the changes caused by attacks. Ishibashi et al.\cite{Ishibashi2010} extracted communication structure to identify low intensity anomalous network events, which can not be detected with conventional volume-based anomaly detection schemes. In a word, the complex network is an effective means for the anomaly detection of structured data. 

\section{Multivariate flow similarity model on temporal locality}
\label{sec:model}
In this paper, multivariate flow similarity on temporal locality is used to quantify the interactive behaviors among similar flows at local time. In the below a specific example of flow interaction is given, i.e., the traffic flow interaction under Google search. During opening the Google.com, the browser firstly looks up the IP address of Google.com by the DNS flows, and loads Google's webpage by HTTP flows. Then, it sends the HTTP request flows again to obtain the search results for the given keywords. Next, we probably click some hyperlinks on search page to trigger new flows. Eventually, we may repeatedly request new flows decided by previous flow content until we obtain our wanted results. Obviously, the flows in each step depend on the previous flows, that is to say the previous flows trigger the further flows. Another example is the recursive or iterative query of DNS resolver. The DNS server has to forward this requests to its provider or tell the user its provider, if there is no record for a given domain name\cite{Mockapetris1988}. In fact, the interactions in large scale traffic flows are so complicated that the traditional statistical methods are very hard to describe the macroscopic relationships. But the complex network has a good nature to depicting the complexity and interaction of large scale traffic flows. Thus, based on complex network theory we propose the construction method of multivariate flow similarity model on temporal locality in the following.

\subsection{Model definition}
\label{subsec:definition}
First, a 6-tuple $f = \{sa, da, sp, dp, pr, ps\}$ is defined to denote one network traffic flow, where the $sa$, $da$, $sp$, $dp$, $pr$, and $ps$ represent the source IP, destination IP, source port, destination port, protocol number, and flow payload size respectively. Then the set $F = \{ f_{1}, f_{2}, \dots, f_{n} \}$ denotes $n$ traffic flow traces on time series. When we focus on traffic flows of one Internet application (e.g., either of two ports is a fixed value), the 6-tuple can be simplified into the 3-tuple $f=\{sa, da, ps\}$. Here one unique flow can be identified by the values of $sa$ and $da$.

\subsubsection{Temporal locality}
Temporal locality is used to describe the interaction on a per item basis apart from in an aggregate reference flow\cite{MAHANTI2000187}. In network traffic flows, the interactive relationships are built from current flows to further flows. For two flows $f_i$ and $f_j$ occurring at time $t_i$ and $t_j$ separately ($t_i \leq t_j$ ), if $t_j \in [t_i, t_i + \Delta w]$, there would be the interactive relationship from $f_i$ to $f_j$, where the $\Delta w$ denotes the size of temporal locality window. In other words, there will be a directed connection from flow node $f_i$ to $f_j$. However, the captured network traffic flows are diverse, especially those at the backbone network. Two flows $f_i$ and $f_j$ could be irrelevant in flow content in despite of $t_j \in [t_i, t_i + \Delta w]$ or $t_i \in [t_j, t_j + \Delta w]$. In order to eliminate the impacts from irrelevant flows, we design the multivariate flow similarity method to filter the pseudo interactive relationships of the flows by quantifying the similar probability of two flows.

\subsubsection{Multivariate flow similarity}
To determine true relationship of two flows, we propose a multivariate flow similarity method by computing the similar probability of features values of two flows, such as the source and destination IP, source and destination port, protocol type, and flow payload size.

\textbf{Source and Destination IP addresses}. 
Source and destination IP addresses have been widely used in the intrusion detection domain. Whereas as a feature it does not provide a definitive conclusion, it can be used as a reference in network traffic flow profiling. Different from statistic-based traffic anomaly detection, in this paper their primary value comes from a similar probability to evaluate the relevance of different flows. For example, the attack flows from single attacker to multiple victims contain the attacker's IP address. Thus, we develope the below equation \eqref{eq:model-ra} to calculate flow similarity on IP address.
\begin{equation}
	r_{a}(f_i, f_j) = \frac{max\{ LCP_1,LCP_2,LCP_3,LCP_4\}}{L}
	\label{eq:model-ra}
\end{equation}
Where the $LCP_1 = |sa_i \cap sa_j|$ , $LCP_2 =|sa_i \cap da_j|$, $LCP_3 = |da_i \cap sa_j|$, $LCP_4 = |da_i \cap da_j|$ denote the length of common prefix of two IP addresses, and the $L$ is the IP address length, i.e. 32 for IPv4 and 128 for IPv6.

\textbf{Source and Destination port}.
In TCP/IP network, a port number is a way to identify a specific application to which an Internet or other network message is to be forwarded when it arrives at a server. So the port number is usually used to match the Internet services of the corresponding TCP or UDP implementation, such as the 21 for FTP, 22 for SSH, 23 for TELNET, 25 for SMTP, 53 for DNS, 80 for HTTP, 6881-6889 for BT and 5554 for Worm Sasser. The flows from same Internet services would like to interact each other. Thus, as shown in equation \eqref{eq:model-rpo}, if existing two ports $p_i$ and $p_j$ in flows $f_i$ and $f_j$ belong to same Internet services, the probability of flow interaction is 1, otherwise 0.
\begin{equation}
	r_{po}(f_i, f_j)=
	\begin{cases}
		1 & if \: p_i \cong p_j, \exists{p_i} \in f_i \: and \: \exists{p_j} \in f_j \\
		0 & otherwise
	\end{cases}
	\label{eq:model-rpo}
\end{equation}

\textbf{Protocol}.
Similar to source-destination port pair, protocol was widely used for filtering out non-related traffic, thus reducing the volume of flows requiring further processing. Sometimes the sheer presence of a specific protocol in traffic raises suspicion. For example, IRC traffic is relatively rarely used for legitimate purposes to the extend that in certain networks there is no use for this protocol at all. So the flows belonging to the same protocol in a monitored network, especially the special protocols, are more probable to interact each other. 
\begin{equation}
	r_{pr}(f_i, f_j)=
	\begin{cases}
		1 & pr_i = pr_j \\
		0 & otherwise
	\end{cases}
	\label{eq:model-rpr}
\end{equation}

\textbf{Flow payload size}.
The flow payload size are mostly intended to represent similar communication patterns. This metric has been used with the purpose of both traffic classification (i.e. distinguish specific protocols, especially the P2P) and traffic anomaly detection. That is because continuous flows from one Internet application exhibit a very consistent behavior. Here we use the ratio between two flows' payload sizes to exhibit the flow similarity.
\begin{equation}
	r_{ps}(f_i, f_j) = \frac{min\{ps_i, ps_j\}}{max\{ps_i, ps_j\}}
	\label{eq:model-rps}
\end{equation}
Considering the normalization of the $r_{ps}$, in equation \eqref{eq:model-rps} the minimum of two flows' payload size is divided by the maximum of that.

Furthermore, the multivariate similarity between two flows can be calculated by the following equation:
\begin{equation}
	\resizebox{0.9\hsize}{!}{$
			r(f_i, f_j) = w_a * r_a + w_{po} * r_{po} + w_{pr} * r_{pr} + w_{ps} * r_{ps},
			\label{eq:model-r}
		$}
\end{equation}
where the $w_a$, $w_{po}$, $w_{pr}$, and $w_{ps}$ denote the weight of corresponding flow feature respectively, $w_a + w_{po} + w_{pr} + w_{ps} =1$. A entropy weight method \cite{Chen201705} is adopted to determine the weights of equation \eqref{eq:model-r}. Furthermore, the interaction relationship of two flow nodes $f_i$ and $f_j$ is expressed as:
\begin{equation}
	E_{ij}=
	\begin{cases}
		1 & t_j \in [t_i, t_i+\Delta w]\;and \; r(f_i,f_j) \geq r_c \\
		0 & otherwise
	\end{cases}
	\label{eq:model-c}
\end{equation}
If $E_{ij} = 1$, an connection is built from flow node $f_i$ to flow node $f_j$, and vice versa. Specifically, if only the similarity $r(f_i, f_j)$ between two flows is not less than a critical threshold $r_c$ and the $f_j$ occurs in the temporal locality window of the $f_i$, there should be a directed connection from $f_i$ to $f_j$. The parameters $\Delta w$ and $r_c$ should be determined to an proper value by which the complex network can capture the characteristics of network traffic time series. We have discussed it in detail in Section \ref{subsec:parameters}.

According to the above method, a complex network model $g=( N, E)$ based on multivariate flow similarity on temporal locality can be constructed from network traffic traces, where the $v \in  N$ is the network node that denotes unique network flow $f$, the $e \in E$ describes a interaction relationship between two nodes. In this paper the $g$ is a directed complex network, where the direction of edge $e(v_i, v_j)$ indicates the flow $v_i$ triggers the flow $v_j$, the $v_j$ depends on the $v_i$, or there is an relevance. About MFS-TL, the most important is the ability of describing macroscopic interactive structure, even though it builds the connections from microscopic traffic flows. Therefore, the MFS-TL can capture the interaction and dynamic of large scale network traffic flows.

\subsection{Model filtration and formation}
\label{subsec:filtration-formation}
One of the fundamental questions in using  MFS-TL is the definitions of network node and edge. This basic question can be answered in many different ways depending on the goal of our study. We start with the observation that what kind or level of network flows should be selected as network node in  MFS-TL. We call this process \textit{Node Filtering}. One simple node filtering is to select IP protocol flows. In addition to this basic node filtering, we can enrich the definition of what constitutes a node by imposing "stricter" rules that capture different aspects of traffic flows. For instance, we can have filters for "allowing" a flow node based on: (a) the frequency of one flow, (b) the type of the flow (e.g., TCP three-way handshake), (c) the application protocol used (TCP, UDP, ICMP etc.), (d) the application based on port number (e.g., Port Number 80 for HTTP, Port Numbers 6881$-$6889 for BT), and finally (e) looking at properties of the flow content, such as payload size or by using deep packet inspection.

Besides basic definition about network edge in Section \ref{subsec:definition} , more rules or features, called \textit{Edge Filtering}, can be put forward to enrich the definition of network edge. In general, the directed edges can be used to identify the indicator of the probability interaction between a pair of flows. Directed edges in a  MFS-TL are very useful in identifying various node behaviors and also in establishing their causal relationship. However, we could choose to consider undirected edges, which will enable us to use the more extensively studied complex network metrics for undirected networks, as discussed in later Section \ref{subsec:statistical-characteristics}. In addition to edge direction, it is also important to define the level of network edge in  MFS-TL. One simple edge filter is to add an edge $e(v_i,v_j)$ between flow node $v_i$ and $v_j$ when $E_{ij} =1$ of equation \eqref{eq:model-c}. Once an edge is added, this filter ignores any flow interaction from $v_i$ to $v_j$. We call this edge filter as the \textit{Unweighted-Edge(UWE)}, and is mainly used to study the interactive process of network flows. However, for the flow interaction behavior, the frequency of edge $e(v_i,v_j)$ is an important indicator in  MFS-TL. We call this edge filter as the \textit{Weighted-Edge(WE)}.

In this paper, we mainly focus on the interactions of the traffic flows based on  TCP and UDP application protocol. In other words, the MFS-TL uses the (c) filtering type (as defined above). Throughout the paper and unless stated otherwise, when the legacy application for a flow uses the TCP or UDP, we use the \textit{WE edge filter} on the corresponding protocol of the flows.

Since we use node filtering by application protocol and edge filtering by the edge frequency, the MFS-TLs capture aspects of any application that uses these protocol. However, application protocol-based filtering is consistent with our use of  MFS-TLs as a monitoring tool. For example, if at some time points network traffic at TCP Port 80 appears significantly different, it could be: (a) a new begin or malicious application tunneling its traffic under that port, or (b) a change in the behavior of the traditional traffic.

\section{Anomaly detection method based on intuitionistic fuzzy set}
\label{sec:method}

For a MFS-TL $g = \{ N, E, C\}$, the $C$ is the collection of complex network characteristic metrics. If it is sampled with a fixed time window $\triangle t$, one MFS-TL sequence $G=\{g_1, g_2, \dots , g_n\}$ will be obtained by extracting MFS-TL from network traffic samples, where the $g_i$ represents the flow interactions of a monitored network at $i$th sample period. Assume that there are $p$ characteristic metrics for the sampling MFS-TL $g_i$. Then in $n$ sampling MFS-TLs, existing a characteristic vector $C=\{c\}_{p\times n}$ denotes $p$ time series of MFS-TL characteristics. Every MFS-TL characteristic will depict the MFS-TL structure from different perspectives. For instance, the number of MFS-TL nodes and the number of MFS-TL edges describe MFS-TL size. When one anomaly activity occurs, a large number of nodes or edges will disappear or appear suddenly in MFS-TL. The MFS-TL diameter denotes the worst communication path length. Under the intentional attacks, MFS-TL diameter will first increases and then decreases quickly\cite{wang201701}. In Section \ref{subsec:statistical-characteristics}, the detailed analysis results denote that the correlation between network characteristic and network states can be classified into tree types: positive correlation, negative correlation and non-correlation, and is non-deterministic in different datasets. Therefore, we put forward the intuitionistic fuzzification method for single MFS-TL characteristic to quantify three non-deterministic correlations.

\subsection{Intuitionistic fuzzy set for single characteristic}
\label{subsec:ifs}
\begin{definition}[Intuitionistic Fuzzy Set, IFS]
	$X$ is a finite universal set, such as the network diameter values ($X = C_{i}$). An intuitionistic fuzzy set $A$ in $X$ is an object having the following form.
	\begin{equation}
		A=\{ < x, \mu_{A}(x), \gamma_{A}(x), \pi_{A}(x)> | x \in X\}
		\label{eq:ifs}
	\end{equation}
	where the $\mu_{A}(x): X \to [0,1]$ defines the degree of membership, and $\gamma_{A}(x): X \to [0,1]$ defines the degree of non-membership of the element $x \in X$ to intuitionistic fuzzy set $A$, with the condition $0 \leq \mu_{A}(x) + \gamma_{A}(x) \leq 1$ for all $x$ in $X$. For each intuitionistic fuzzy set in $X$, $\pi_{A}(x)=  1 - \mu_{A_j} - \gamma_{A_j}$ is called the hesitation degree(or intuitionistic index) of $x$ to $A$.
\end{definition}

In order to obtain the IFS $A$ of MFS-TL characteristic, we define the domain of discourse, and then partition it. First, the domain of discourse $D = [x_{min} - \varepsilon_1, x_{max} + \varepsilon_2]$ is constructed, where $x_{min}$ and $x_{max}$ are the minimum and maximum of set $X$, and $\varepsilon_1$ and $\varepsilon_2$ are proper positive numbers. Second, the intuitionistic fuzzy C-means clustering algorithm(IFCM)\cite{CHAIRA2011} is used to partition the domain of discourse into $m$ clustering intervals. Meanwhile we get the clustering center vector $V = \{v_1, v_2, \dots, v_c\}$ of universe $D$. Then let
\begin{equation}
	d_{i} = \left \{
	\begin{aligned}
		 & x_{min} - \varepsilon_1, \quad i = 0            \\
		 & (v_{i} +  v_{i+1})/2, \quad i = 1, 2, ..., m -1 \\
		 & x_{max} + \varepsilon_2, \quad i = m            \\
	\end{aligned}
	\right.
	\label{eq:d-func}
\end{equation}
As a result, the universe $D$ is divided into $m$ unequal intervals, i.e. $D = \{ [d_0, d_1], [d_1, d_2],\dots, [d_{m-1}, d_m]\}$. Every $x \in X$ should exist $m$ intuitionistic fuzzy sets $A_i= \{ < x, u_{A_i(x)}, \gamma_{A_i}(x), \pi_{A_i}(x)|x \in D_i\}$, where the $u_{A_i}(x)$ denotes the membership degree of $x$ in $i$th clustering interval $[d_{i-1}, d_i]$ and $\gamma_{A_i}(x)$ is the non-membership degree of that.

Existed methods of membership and non-membership function usually give the hesitation degree a fixed value. Objectively, the hesitation degree should be dynamic with the universal set $X$. So we adopted the Gaussian function (equation \eqref{eq:u-func}) which meets below condition: when the distance between the $x$ and the interval center $v$ is lower, the degree of membership $\mu$ is more close to 1.
\begin{equation}
	u_{A_{i}}(x) = exp(-\frac{(x-\psi_{u_i})^2}{2\sigma_{u_i}^2})
	\label{eq:u-func}
\end{equation}
Where $i = 1, 2, \dots, m$, and $\psi_{u_i}$ and $\sigma_{u_i}$ are the function parameters. Then the following rules are defined so as to resolve above parameters:
\begin{enumerate}
	\item If $x$ is in the middle of an clustering interval, i.e. $x = v_i$, the membership value $\mu_{A_i}(x)$ = 1.
	\item If $x$ is on the boundaries of an clustering interval, i.e. $x = (v_i - v_{i-1} )/2$, let $\pi_{A_i}(x) = \alpha, (0 \leq \alpha \leq 1)$, then $\mu_{A_i}(x) = (1 - \alpha) / 2$.
\end{enumerate}
Based on above rules, the function parameters are resolved:
\begin{align}
	\psi_{u_i}   & = v_i \label{eq:u-func-psi}                                               \\
	\sigma_{u_i} & = -\frac{(v_{i -1} + v_{i})^2}{8ln((1-\alpha)/2)} \label{eq:u-func-sigma}
\end{align}
Thus, given a value $x$, the membership values for every clustering interval are calculated by equations\eqref{eq:u-func}, \eqref{eq:u-func-psi} and \eqref{eq:u-func-sigma}. The non-membership function is calculated based on Yager generating function\cite{BURILLO1996305}. The Yager's intuitionistic fuzzy complement is written as following:
\begin{equation}
	\gamma_{A_i}(x) = (1 - \mu_{A_i}^\beta(x))^{1/\beta}, \quad \beta > 0
	\label{eq:gamma-func}
\end{equation}
When $\mu_{A_i}(x) = 1$, then $\gamma_{A_i}(x) = 0$, and otherwise vice versa. Therefore the IFS (equation\eqref{eq:ifs}) becomes:
\begin{equation}
	\begin{split}
		A_i= & \{ < x, \mu_{A_i}(x), (1 - \mu_{A_i}^\beta(x))^{1/\beta}, \\
		& 1- \mu_{A_i}(x) - (1 -  \mu_{A_i}^\beta(x))^{1/\beta}> | x \in D_i\}
		\label{eq:ifs-2}
	\end{split}
\end{equation}

As we all known, one network is either normal or abnormal. But the value of clustering intervals $m$ is not less than 2 depending on our goals. When $m=2$, one clustering interval represents normal state and the other represents abnormal state. When $m > 2$, as Fig.\ref{fig:a-cluster} shown the anomaly detection performance achieves the best when $m = 10$, how to partition $m$ clustering intervals again to two states' clustering interval set becomes an important problem for anomaly detection. In other words, we need to know which clustering intervals represent the normal state and the others represent the abnormal state. Let the $NC$ and $AC$ denote the normal clustering interval set and abnormal clustering interval set, $D = AC \bigcup NC$ and $AC \bigcap NC = \emptyset$. The abnormal clustering interval set $AC$ is that includes most of the abnormal instances of the training set, and the normal clustering interval set $NC$ is that includes most of the normal instances of the training set. For the $i$th abnormal clustering interval in $AC$, the $tt_i = \frac{AI_i}{|AC_i|} $ and $tf_i=\frac{NI_i}{|AC_i|}$ denote the ratio of the number of abnormal instances $AI_i$ and the number of normal instances $NI_i$ to total instances $|AC_i|$. Correspondingly, the $ff_j = \frac{NI_j}{|NC_j|} $ and $ft_j=\frac{AI_j}{|NC_j|}$ denote the ratio of the number of normal instances $NI_j$ and the number of abnormal instances $AI_j$ to total instances $|NC_j|$ of the $j$th normal clustering interval. For the two states' set, we get $TT = \sum_{i=0}^{i\in AC}tt_i$, $ TF = \sum_{i=0}^{i\in AC}tf_i $, $FF = \sum_{j=0}^{j\in NC}ff_j$ and $FT = \sum_{j=0}^{j\in NC}ft_j$. Furthermore, a distinction index is designed for the purpose that finds the best partitioning for the clustering intervals by obtaining the maximum $\eta = TT - TF + FF-FT$.
\begin{equation}
	\tau = \frac{\eta}{TT +TF + FF+FT}, \argmax_{AC, NC}(\eta)
	\label{eq:dist-func}
\end{equation}
In above equation, we found that the best case is there is no normal instances in abnormal clustering interval set and no abnormal instances in normal clustering interval set, i.e. $TT=1$, $TF=0$, $FF=1$, and $FT=0$. When all instances of training set distribute evenly, the $TT$, $TF$, $FF$, and $FT$ will be 0. Thus, the $\tau \in [0,1]$ denotes distinction degree of the clustering intervals to network states. The higher the $\tau$ is, the better the clustering intervals' partition is, and the more significant the relevance of MFS-TL characteristic values to network states is. Essentially, it is because the changes of the characteristic values can reflect network states.

Here we find that the $A$, as the individual detector for single MFS-TL characteristic, can give detection result by selecting a IFS with maximum membership degree.

\subsection{Intuitionistic fuzzy set ensemble for multiple characteristics}
\label{subsec:ifse}
Single MFS-TL characteristic as the individual detector alone would be used to detect abnormal in some datasets. But the studies of complex network anomaly detection indicate that the changes of each of characteristics have the inconsistent performance in same anomaly events \cite{Ai2013,wang201701,Rayana2016}. So a ensemble method of multiple characteristic IFSs is essential for the purpose of eliminating the inconsistent.

For the temporal sequence of one MFS-TL characteristic $C_i$, we can compute its domain of discourse $D_i$ and its IFS $A_i$ based on Section \ref{subsec:ifs}. Furthermore equation\eqref{eq:ifs-2} can be extended as following equation for multiple MFS-TL characteristics:
\begin{equation}
	A_{ij}(c)=\{<c,\mu_{A_{ij}}(c),\gamma_{A_{ij}}(c),\pi_{A_{ij}}(c)>| c \in D_j\}
	\label{eq:ifs-3}
\end{equation}
Where the $i = 1, 2, \dots, p$, $j = 1, 2, \dots, m$, and the $D_j$ is the $j$th clustering interval of one MFS-TL characteristic $C_i$. The $A_{ij}(c)$ represents the IFS of the $i$th network characteristic $c$ to the $j$th clustering interval. In other words, the $\mu_{A_{ij}}(c)$ is the membership function of $i$th network characteristic value to $j$th clustering interval of network characteristic universe $D_i$, the $\gamma_{A_{ij}}(c)$ is the non-membership function of $i$th network characteristic value to $j$th clustering interval of the universe $D_i$, and the $\pi_{A_{ij}}(c)$ is the hesitation degree.

Finally, the intuitionistic fuzzy relationship $\boldsymbol{A}$ between $p$ MFS-TL characteristics and $m$ clustering intervals are calculated by carrying out the temporal sequence partition and intuitionistic fuzzy set construction on the training set. Hence, the above problem becomes the multi-IFSs reasoning problem.  Let $\boldsymbol{C}' =[c_1\quad c_2 \dots c_p]^{T}$ denotes the characteristic values of a testing MFS-TL $g'$. Then we define the equation \eqref{eq:ifs-4} to compute the IFSs $\boldsymbol{B}$ of the characteristic collection $\boldsymbol{C}'$ to $m$ clustering intervals.
\begin{equation}
	\begin{aligned}
		\boldsymbol{B}=A\odot \boldsymbol{C'}=\left [
			\begin{aligned}
				 & A_{11}(c_1)  & \hspace{-0.8em} A_{12}(c_1)  & \dots\hspace{-0.8em}  & A_{1m}(c_1)   \\
				 & A_{21}(c_2)  & \hspace{-0.8em} A_{22}(c_2)  & \dots\hspace{-0.8em}  & A_{2m}(c_2)   \\
				 & \quad \vdots & \hspace{-0.8em} \vdots\qquad & \vdots\hspace{-0.8em} & \vdots \qquad \\
				 & A_{p1}(c_p)  & \hspace{-0.8em}  A_{p2}(c_p) & \dots\hspace{-0.8em}  & A_{pm}(c_p)   \\
			\end{aligned}
			\right ]^{T}
	\end{aligned}
	\label{eq:ifs-4}
\end{equation}
Where the $B_{ij}$ denotes the membership, non-membership and hesitation of the $j$th characteristic value to the $i$ clustering interval, and then the $B_i$ is the IFS that current network depicted by $p$ network characteristics is mapped to the $i$th clustering interval. As a result, the $\boldsymbol{B}$ describes the intuitionistic fuzzy set between the network characteristics and clustering intervals. 

However, for multiple characteristic IFSs $\boldsymbol{B}$, the sizes of normal interval set $NC$ and abnormal interval set $AC$ are usually different with different MFS-TL characteristic, e.g. the size of the $AC$ of network edge $|AC^{edge}|$ may be not equal that of network diameter $|AC^{diameter}|$. Even if their sizes are equal, the items of the set $NC$ and $AC$ of each MFS-TL characteristic may not be one-to-one correspondence to the domain of discourse $D$, e.g. $i \neq j$ where $AC_0^{edge}=D_i^{edge}$ and $AC_0^{diameter} = D_j^{diameter}$. In order to fuse multiple characteristics IFSs, we introduce the intuitionistic fuzzy weighted average operator to separately combine the intuitionistic fuzzy sets in clustering interval set $NC$ and $AC$. Furthermore, the IFSs $\boldsymbol{B}$ between $p$ network characteristics and $m$ cluster intervals are translated into the IFSs $\boldsymbol{B'}$ with $p$ network characteristics and two states' linguistic variables $L=\{L_1, L_2\}$ based on the distinction index function \eqref{eq:dist-func} and intuitionistic fuzzy weighted average operator.
\begin{equation}
	\begin{aligned}
		\boldsymbol{B'}= \left [
			\begin{aligned}
				 & A_{11}'(c_1) & \hspace{-0.8em} A_{12}'(c_1)  \\
				 & A_{21}'(c_2) & \hspace{-0.8em} A_{22}'(c_2)  \\
				 & \quad \vdots & \hspace{-0.8em} \vdots\qquad  \\
				 & A_{p1}'(c_p) & \hspace{-0.8em}  A_{p2}'(c_p) \\
			\end{aligned}
			\right ]^{T}
	\end{aligned}
	\label{eq:ifs-5}
\end{equation}
Where $A_{i1}'(c_i)=\{<c_i,\mu_{A_{i1}'}(c_i),\gamma_{A_{i1}'}(c_i),\pi_{A_{i1}'}(c_i)>| c_i \in L_1\}$ and $A_{i2}'(c_i)=\{<c_i,\mu_{A_{i2}'}(c_i),\gamma_{A_{i2}'}(c_i),\pi_{A_{i2}'}(c_i)>| c_i \in L_2\}$, and the $L_1$ and $L_2$ denote the abnormal clustering interval set $AC$ and the normal clustering interval set $NC$ separately. In order to judgment whether the network is abnormal or not based on multi-characteristic IFSs $\boldsymbol{B'}$, the IFWG is introduced to fuse IFSs of multiple network characteristics to the linguistic variables.
\begin{definition}[IFWG]
	Let $\alpha_j=(\mu_j, \gamma_j)$ be a collection of IFS with the weight vector $w = (w_1, w_2, \dots, w_p)^{T}$ such that $w_j \in [0,1]$ and $\sum_{j}^{p}{w_j} =1$. An IFWG operator of dimension $p$ is a mapping: $\omega^{p} \to \omega$, and
	\begin{equation}
		\begin{split}
			IFWG_{\omega}(\alpha1, \alpha2,...,\alpha_{p}) = \alpha_1^{\omega_1}\oplus\alpha_2^{\omega_2}\oplus...\oplus\alpha_{p}^{\omega_p} \\
			= (\prod_{j=1}^{p}u_{a_j}^{\omega_j}, 1- \prod_{j=1}^{p}(1 - v_{a_j})^{\omega_j}))
		\end{split}
		\label{eq:ifwg}
	\end{equation}
\end{definition}
Moreover, the weight $w_i$ of the $i$th network characteristic $c$ is calculated based on the distinction index $\tau$ of MFS-TL characteristic, as the following equation \eqref{eq:ife-w}:
\begin{equation}
	w_i = \frac{\tau_i}{\sum_{i=1}^{p}(\tau_i)}.
	\label{eq:ife-w}
\end{equation}
Using the IFWG, the IFS $B''$ of MFS-TL structure to each of linguistic variables is calculated as shown in equation\eqref{eq:ifwg-2}. Where the $\mu_{B'_i}(C')$ is the degree of membership of the network structure to the $i$th linguistic variable, the $\gamma_{B'_i}(C')$ and $\pi_{B'_i}(C')$ are the non-membership and hesitation of that. Then, we defined a method of comparing multiple IFSs, based on the score function and the precision function, to obtain the defuzzification of the $B'$.
\begin{equation}
	\begin{aligned}
		B''
		 & = IFWG_{\omega}(B') \\
		 & =\left (
		\begin{aligned}
				 & \{<C', \mu(C'), \gamma(C'), \pi(C')> | C' \in L_1 \} \\
				 & \{<C', \mu(C'), \gamma(C'), \pi(C')> | C' \in L_2 \} \\
			\end{aligned}
		\right )
	\end{aligned}
	\label{eq:ifwg-2}
\end{equation}
\begin{definition}[Score function\cite{CHEN1994163}]
	For any intuitionistic fuzzy set $A=<\mu, \gamma>$, the score function $S(A)$ of this IFS is defined as follows:
	\begin{equation}
		S(A) = \mu - \gamma, \quad S(A) \in [-1, 1]
		\label{eq:s-func}
	\end{equation}
\end{definition}
It can be seen that the larger the value of $S(A)$ is, the better membership relationship the intuitionistic fuzzy set $A$ is.

\begin{definition}[Precision function\cite{HONG2000103}]
	For any intuitionistic fuzzy set $A = <\mu, \gamma>$, the precision function $H(A)$ of this IFS is defined as follows:
	\begin{equation}
		H(A) = \mu + \gamma, \quad H(A) \in [0, 1]
		\label{eq:h-func}
	\end{equation}
\end{definition}
The larger the value of $H(A)$ is, the higher the precision degree of the intuitionistic fuzzy set $A=<\mu, \gamma>$ is. For any tow intuitionistic fuzzy set $A_1 = <\mu, \gamma>$ and $B = <\mu, \gamma>$, the followings hold true:
\begin{enumerate}
	\item
	      If $S(A) < S(B)$, then $A < B$
	\item
	      If $S(A) = S(B)$:
	      \begin{enumerate}
		      \item
		            When $H(A) = H(B)$, then $A= B$, that is $A$ and $B$ represent the same information.
		      \item
		            When $H(A) < H(B)$, then $A < B$.
	      \end{enumerate}
\end{enumerate}
Based on above rules, the two intuitionistic fuzz logic of the $B''$ can be compared each other. Then the sorted vector is given $R=\{B''^{1}, B''^{2}\}$, and the $B''^{1}$ is the detection result of our method.

\section{Experimental results} 
\label{sec:results}

In order to analyze the complex network feature of our model and verify the anomaly detection performance of our method, we use a variety of publicly available real Internet traffic traces: (1) ISOT dataset,  (2) CTU dataset, and (3) CICIDS2017 dataset. These traces are non-sampled and include up to layer-4 headers with no payload. In this section, our works consist of the analysis of the MFS-TL parameters and MFS-TL statistical characteristics, and the evaluation of our anomaly detection method.

\subsection{Dataset} 
\label{subsec:dataset}

\begin{table*}[htbp]
	\begin{center}
		\caption{Publicly available network traffic datasets from ISOT, CTU and CICIDS project.}\label{tab:dataset}
		\resizebox{1\textwidth}{!}{
			\begin{tabular}{ l  c  c  r  r  r r r r r}
				\hline
				Name       & Date/Time           & Duration & Unique IPs & 2-tuple flows & 5-tuple flows & Abnormal 5-tuple flows & Packets    & Abnormal packets \\
				\hline
				ISOT-06    & 2005-01-06/20:00:00 & 4.33 hrs & 14,492     & 35,362        & 142,955       & 32,945(23.05\%)        & 8,946,619  & 370,099(4.41\%)  \\
				ISOT-07    & 2005-01-07/20:00:00 & 4.00 hrs & 13,906     & 31,269        & 137,307       & 32,945(23.99\%)        & 10,001,278 & 370,099(3.70\%)  \\
				CTU-4      & 2011-08-15/11:00:05 & 4.17 hrs & 185,372    & 211,050       & 722,810       & 1,735(0.24\%)          & 4,171,952  & 3,718(0.09\%)    \\
				CTU-9      & 2011-08-17/12:01:01 & 5.18 hrs & 366,096    & 455,340       & 1,178,257     & 93,736(7.96\%)         & 8,003,477  & 382,708(4.78\%)  \\
				CICIDS-SSH & 2017-07-04/14:00:00 & 2.00 hrs & 4,072      & 7,130         & 66,307        & 2,551(3.85\%)          & 1,703,856  & 5,063(0.30\%)    \\
				CICIDS-DoS & 2017-07-05/09:47:00 & 3.86 hrs & 6,658      & 12,870        & 169,144       & 24,357(14.40\%)        & 6,904,454  & 294,218(4.26\%)  \\
				\hline
			\end{tabular}
		}
	\end{center}
\end{table*}

The ISOT dataset is the combination of several existing publicly available malicious and non-malicious datasets, such as the abnormal dataset from the French chapter of the honeynet project, and the normal datasets from the Traffic Lab at Ericssion Research in Hungary and the Lawrence Berkeley National Lab (LBNL) . As mentioned in Reference \cite{Saad2011}, the experimental data is produced by merging multiple original datasets into a single individual trace file. Based on this combined traces, we selected two pieces of data: the LBNL normal data on Jan 6, 2005 and Jan 7, 2005, and the abnormal data captured at 22:00$-$23:00 on Oct 7, 2010. Then, having the normal traffic at 2005-01-06 20:00:00$-$2005-01-07 00:26:00 and 2005-01-07 20:00:00$-$2005-01-08 00:26:00 as background traffic, the abnormal traffic is separately injected into background traffic. Furthermore, we obtain the experimental datasets ISOT-06 and ISOT-07 which include the Storm and Waledac anomalous activities.

The CTU is a dataset of anomalous traffic that was captured in the CTU University, in 2011. The goal of the dataset was to have a large capture of real botnet traffic mixed with normal traffic. The CTU dataset consists in thirteen captures of different botnet samples. On each sample a specific malware was executed, which used several protocols and performed different actions. In this paper, the samples CTU-4 and CTU-9 are adopted which executed the malicious softwares Rbot and Neris, respectively\cite{GARCIA2014}.

The CICIDS2017 dataset contains benign and the most up-to-date common attacks, which resembles the true real-world data\cite{Massicotte2006}. The data capturing period started at 9:00, July 3, 2017 and ended at 17:00 on July 7, 2017, for a total of 5 days. The first day is the normal day and only includes the benign traffic. The implemented attacks include Brute Force FTP, Brute Force SSH, DoS, Heartbleed, Web Attack, Infiltration, Botnet, and DDoS. They have been executed both morning and afternoon on the other days. The SSH-Patator(i.e. Brute Force SSH attack) and CICIDS-DoS (including the DoS-Slowloris, DoS-Slowhttptest, DoS-Hulk, DoS-GoldenEye, and DDoS-LOIT attacks) are selected in our work. Moreover, the normal traffic from the fist traces with corresponding time period of the attack event is also extracted as the normal traffic of this attack. 

Table \ref{tab:dataset} summarizes the captured time, the captured duration, and the number of unique IPs, 2-tuple flows ($f=\{sa, da\}$), 5-tuple flows ($f=\{sa, da, sp, dp, pr\}$), abnormal 5-tuple flows, network packets and abnormal packets. Noted that the values in parentheses denote the percentage of the abnormal flows. Based on existed data labels, we propose a new label method because the MFS-TL is constructed on the sampled traffic traces with a fixed time interval. Specifically, we calculate the ratio $\gamma$ between the number of abnormal flows and the total flows in one sampled traffic traces. If the ratio $\gamma$ is not less than $0.001$, this sample is labeled with abnormal, otherwise it is normal.

Additionally, we have used four following common evaluation metrics: accuracy, precision, recall, and F-Measure. Accuracy ($Acc$) is the percentage of all normal and anomaly instances that are correctly classified. Precision ($Pre$) is the percentage of correctly detected anomaly instances over all detected anomaly instances. Recall ($Rec$) is the percentage of anomaly instances correctly detected. F-Measure ($F_1$) is a harmonic combination of the precision $Pre$ and recall $Rec$ into a single measure. Based on above definitions, we give following equations:
\begin{equation}
	\begin{aligned}
		Acc & = \frac{TP + TN}{TP + TN + FP + FN} \\
		Pre & = \frac{TP}{TP+FP}                  \\
		Rec & = \frac{TP}{TP+FN}                  \\
		F_1 & = \frac{2*Pre*Rec}{Pre + Rec}       \\
	\end{aligned}
\end{equation}
Where the True Positives ($TP$) measures how many instances of a given class are correctly classified; the True Negatives ($TN$) measures the number of correctly classified instances of a class; the False Positives ($FP$) measures how many instances of other classes are confused with a given class; and the False Negatives ($FN$) measures the number of misclassified instances of a class. 

\subsection{Parameter analysis of the MFS-TL}
\label{subsec:parameters}
\begin{figure}[ht]
	\begin{center}
		\captionsetup[subfloat]{labelformat=empty,farskip=2pt,captionskip=1pt}
		\subfloat{%
			\includegraphics[width=0.495\columnwidth]{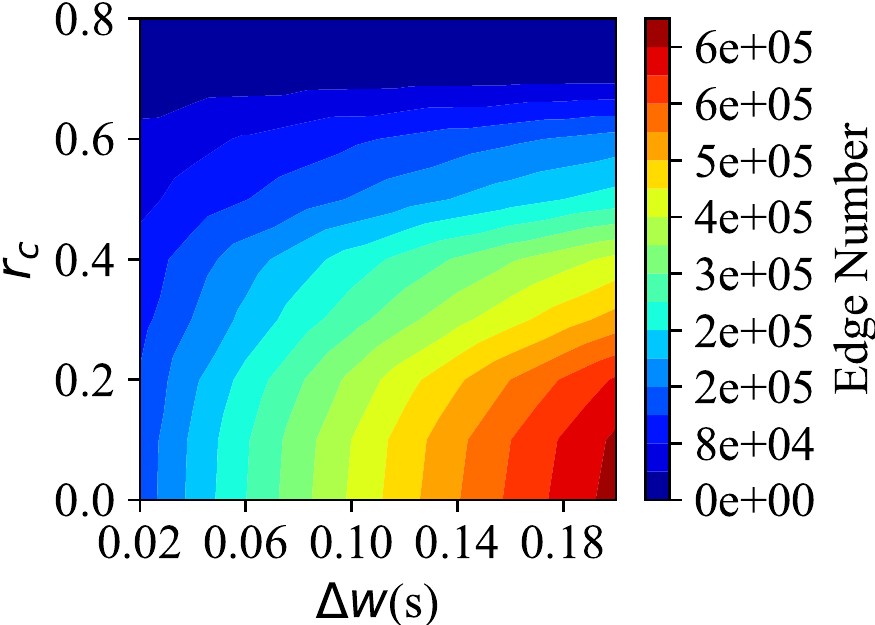}%
			\label{sfig:params-ctu-9-edge}
		}
		\subfloat{%
			\includegraphics[width=0.495\columnwidth]{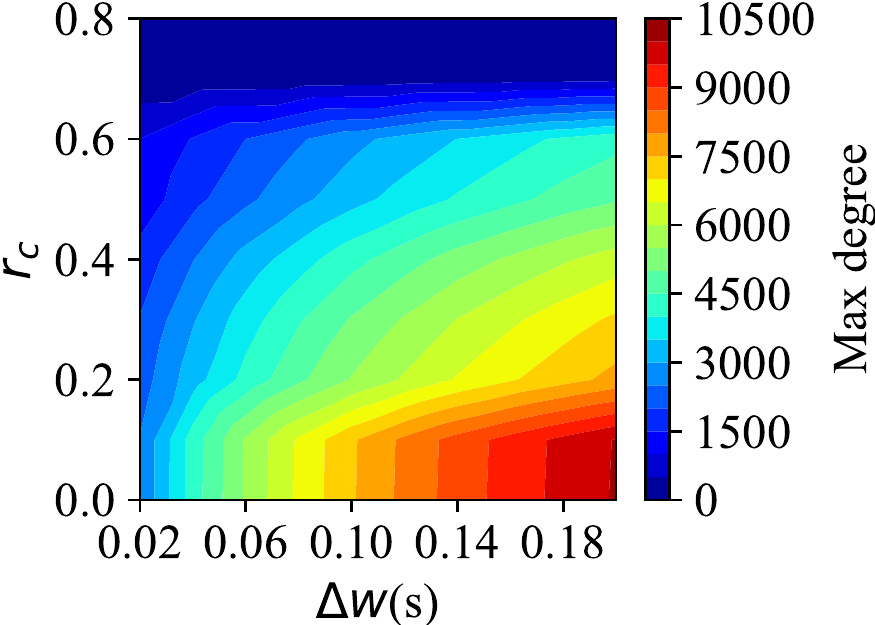}%
			\label{sfig:params-ctu-9-degree-max}
		}\\
		\subfloat{%
			\includegraphics[width=0.495\columnwidth]{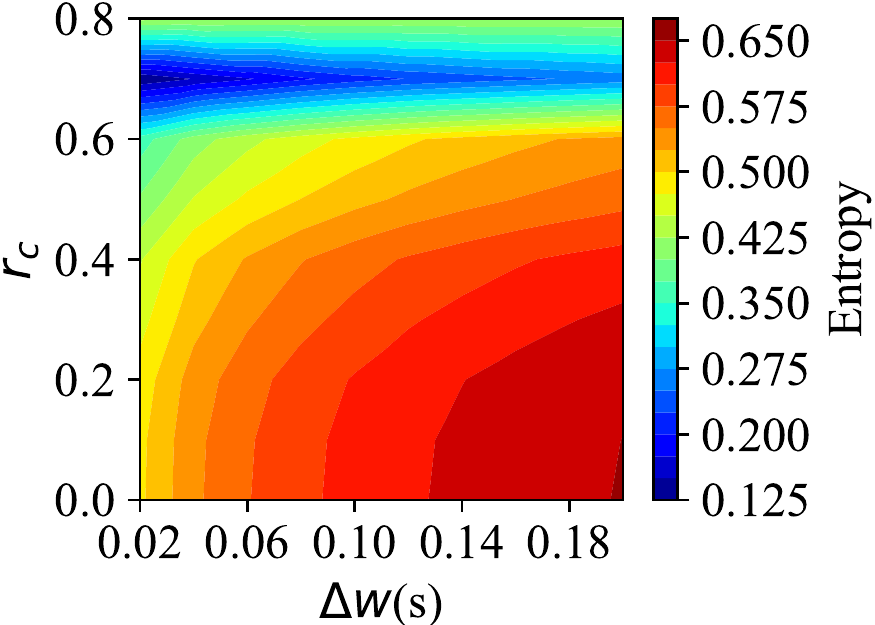}%
			\label{sfig:params-ctu-9-entropy}
		}
		\subfloat{%
			\includegraphics[width=0.495\columnwidth]{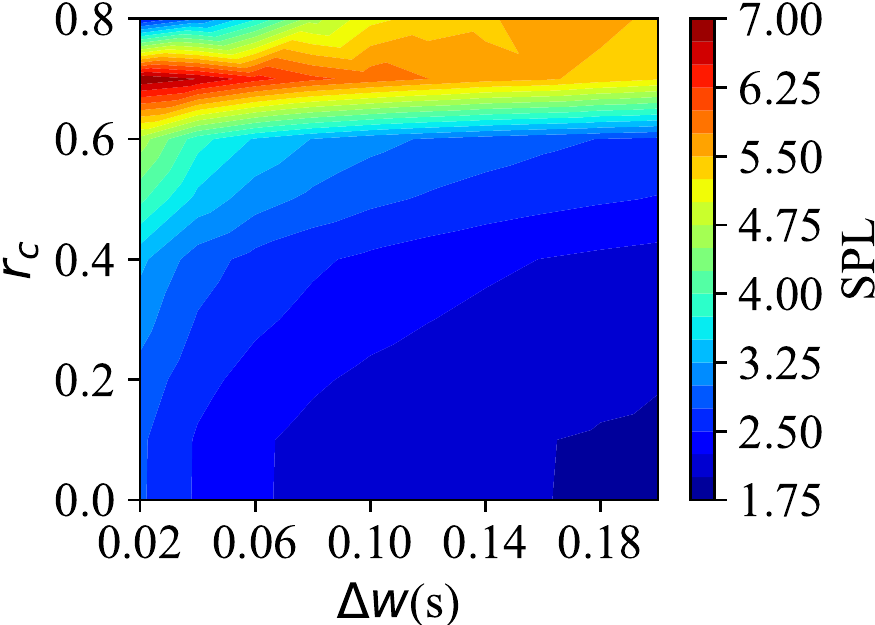}%
			\label{sfig:params-ctu-9-spl}
		}
		\caption{Distributions between temporal locality window ($\Delta w$), multivariate flow similarity critical threshold $r_c$ and MFS-TL characteristics values in CTU-9 dataset.}
		\label{fig:params-ctu-9}
	\end{center}
\end{figure}

As defined in MFS-TL model, the temporal locality window $\Delta w$ and multivariate flow similarity critical threshold $r_c$ are two key parameters which determine the connectivity among the flows. Thus how to select an appropriate temporal locality window and multivariate flow similarity critical threshold are very important for MFS-TL structure. Theoretically, the bigger $\Delta w$ or lower $r_c$ offers the higher connection probability of two flow nodes. As a result, the network density of the MFS-TL increases with the above parameter values. However, if the $\Delta w$ is big enough that the MFS-TL loses the purpose of depicting the interaction relationship among the flows. When the $r_c$ is close to 0, every flow becomes the isolated node of the MFS-TL. Conversely, if the $\Delta w$ is too small or the $r_c$ is too high, some valuable connections in MFS-TL would be filtered. Thus the $\Delta w$ and $r_c$ should be determined to the proper values by which the complex network MFS-TL can capture the interactive feature of the network traffic flow time series. For the purpose of constructing a proper MFS-TL, a group of complex network characteristics are analyzed as a function of the $\Delta w$ and $r_c$, as shown in Fig.\ref{fig:params-ctu-9}.

First, the network traffic traces from the ISOT-06, CTU-9, and CICIDS-SSH are split with the 1 minute sampling time. Second, given the $\Delta w$ and $r_c$, we can construct one MFS-TL for one sampled traffic traces. Thus, each of traffic datasets is translated into the MFS-TL sequences with the fixed $\Delta w$ and $r_c$. Then, we calculate the characteristics of all MFS-TLs, such as the node number, edge number, mean degree, max degree, clique, K-core, entropy, clustering coefficient, assortative coefficient, SPL, diameter(mean), diameter(max) and power-law of network degree distribution, as shown in Table \ref{tab:characteristic}. Finally, the contour is used to plot the distributions of the characteristic values as a function of the $\Delta w$ and $r_c$. Due to the space limitation, only four distributions including the edge number, mean degree, SPL, and entropy of the CTU-9 are presented in Fig.\ref{fig:params-ctu-9}. The statistical results from ISOT-06, CTU-9, and CICIDS-SSH show that the distributions about mean degree, clique, and K-core are similar with the Fig.\ref{sfig:params-ctu-9-edge} and \ref{sfig:params-ctu-9-degree-max}, the assortative, SPL, and diameter(max) are similar with the Fig.\ref{sfig:params-ctu-9-spl}. But the distribution of the clustering or power-law is different from the other characteristics, and is different with different datasets.
\begin{table}[htbp]
	\begin{center}
		\caption{The used characteristic metrics for the MFS-TL.}\label{tab:characteristic}
		\resizebox{1\columnwidth}{!}{
			\begin{tabular}{| p{0.20\columnwidth} | p{0.75\columnwidth} |}
				\hline
				Characteristic & Description                                             \\
				\hline
				Node number    & The number of network nodes.                            \\
				Edge number    & The number of network edges.                            \\
				Mean Degree    & The average degree connectivity of a network.           \\
				Max Degree     & The maximum degree connectivity of a network.           \\
				MDR            & The ratio of the maximum degree in a network.           \\
				K-core         & The maximum sub-network in which node degree $\geq k$.  \\
				Clique         & The number of the largest clique of a network.          \\
				Clustering     & The transitivity of all nodes in a network.             \\
				Assortative    & A preference for a network's nodes to attach to others. \\
				Entropy        & The measure of unpredictability of network structure.   \\
				SPL            & The average of shortest paths of all pairs of nodes.    \\
				Diameter(max)  & The maximum of shortest paths of all pairs of nodes.    \\
				Diameter(mean) & The maximum of average paths of all pairs of nodes.     \\
				Power-law      & The slope of best-fit line for degree distribution.     \\
				\hline
			\end{tabular}
		}
	\end{center}
\end{table}

Observing the distributions of Fig.\ref{fig:params-ctu-9}, we found that the values of edge or max degree increase linearly, the entropy grows as a power of the $\Delta w$, and the SPL decreases as a power of the $\Delta w$, when the $r_c$ is a fixed value. However, when the $\Delta w$ is a fixed value, the values of edge, max degree, and entropy decrease exponentially, and the SPL grows exponentially. But, when the $ 0.6 < r_c < 0.8$, the changes of the edge number and the max degree become relatively slow. However, the entropy and SPL have mutation correspondingly. As we all know, the lower the systemic entropy is, the easier it is to accurately describe its microscopic state\cite{Crooks1999}. Thus, in this paper the multivariate flow similarity critical threshold $r_c$ is selected at the mutation point, i.e. $r_c = 0.65$. Meanwhile considering the IP latency, that the communication latency within a country is usually 0.03s$-$0.05s and that between two countries, especially the transcontinental, is 0.1s$-$0.15s \footnote{\url{ http://www.verizonenterprise.com/about/network/latency/}}, the temporal locality window $\Delta w$ is set as 0.1s. 

\subsection{Statistical characteristics of the MFS-TL}
\label{subsec:statistical-characteristics}

\begin{figure}[ht]
	\begin{center}
		\captionsetup[subfloat]{labelformat=empty,farskip=2pt,captionskip=1pt}
		\subfloat{%
			\includegraphics[width=0.80\columnwidth]{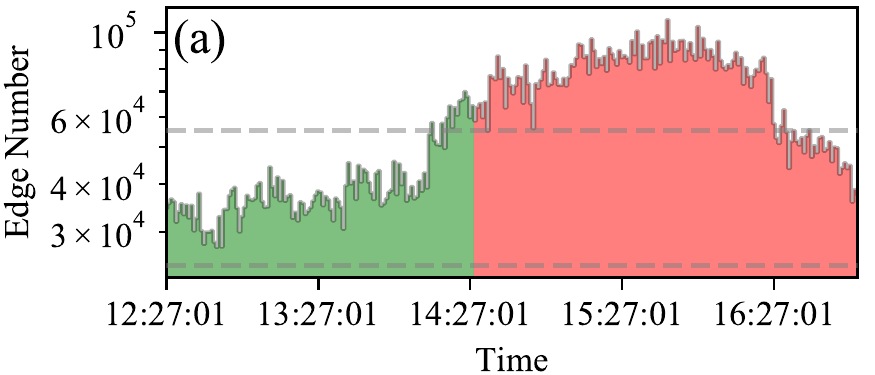}%
			\label{sfig:ch-ctu-9-edge}
		}\\
		\subfloat{%
			\includegraphics[width=0.80\columnwidth]{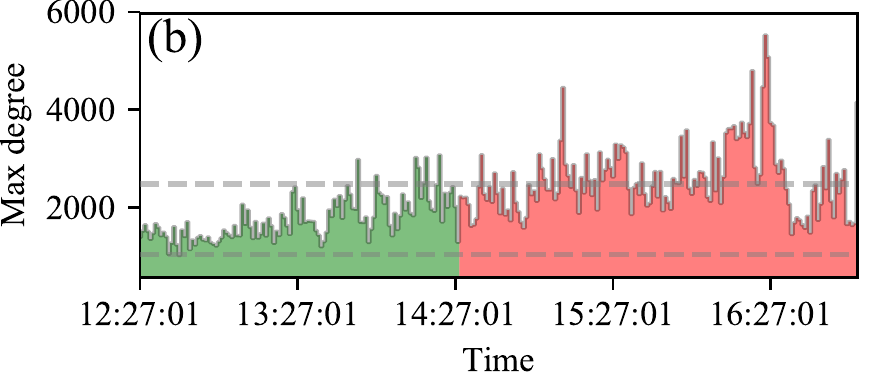}
			\label{sfig:ch-ctu-9-degree-max}
		}\\
		\subfloat{%
			\includegraphics[width=0.80\columnwidth]{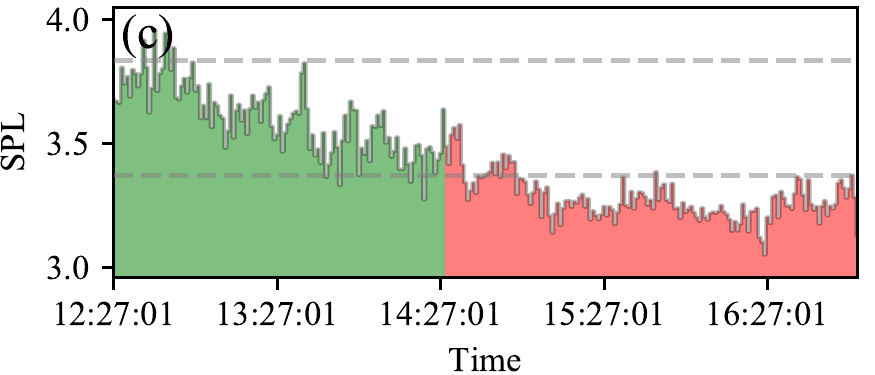}%
			\label{sfig:ch-ctu-9-spl}
		}\\
		\subfloat{%
			\includegraphics[width=0.80\columnwidth]{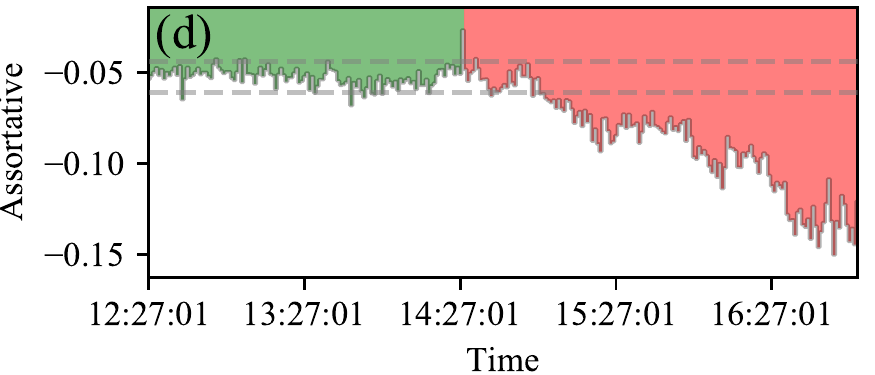}%
			\label{sfig:ch-ctu-9-assortative}
		}
		\caption{The characteristic evolution of MFS-TLs from the CTU-9 dataset. (a) edge number vs. time, (b) max degree vs. time, (c) shortest path length (SPL) vs. time, and (d) assortative vs. time. The left green area and right red area of each sub-figure denote normal traffic and attack traffic. The dashed lines of each of sub-figures represent the upper and lower of the confidence interval with the confidence level $1-\alpha = 0.9$}
		\label{fig:ch-ctu-9}
	\end{center}
\end{figure}

In order to describe the structure feature of MFS-TL, 14 network characteristic metrics are introduced in Table \ref{tab:characteristic}. Each of MFS-TL characteristic metrics will present specific structure feature depending on itself definition. For instance, the node number and edge number denote the network size. The MDR shows the importance of max-degree node in a network. The SPL and the diameter can evaluate the transmission performance of a network. The power-law points out the node preference attachment in a network. In this section, the CTU-9 dataset is used to study the dynamical evolution of the MFS-TLs with anomaly events. Based on above method, the sampling MFS-TL was constructed, the MFS-TL characteristics were calculated, and then the characteristic values was plotted as a function of time. But due to the space limitation, in Fig.\ref{fig:ch-ctu-9} we only exhibit the dynamic evolution of some characteristics including the edge number, max degree, SPL and assortative coefficient. Following the method proposed in Section \ref{subsec:dataset}, the real network states (i.e., normal or abnormal) are labeled over time tick. In Fig.\ref{fig:ch-ctu-9}, the left green area and right red area of each sub-figure denote normal traffic and attack traffic. Clearly, the correlation between the network characteristic evolution and network states can be classified into three types: positive correlation, negative correlation and non-correlation. For instance, in the CTU-9 dataset the evolutions of the node number, edge number, mean degree, clustering, K-core, clique and entropy are positive correlative with network states (e.g. Fig.\ref{sfig:ch-ctu-9-edge}), those of the SPL, diameter(mean), and diameter(max) are negative correlative with network states (e.g. Fig.\ref{sfig:ch-ctu-9-spl} and \ref{sfig:ch-ctu-9-assortative}), and the MDR, max degree, power-law are non-correlative with network states (e.g. Fig.\ref{sfig:ch-ctu-9-degree-max}). It suggests that the MFS-TL characteristic measurement will be a effective method for Internet traffic anomaly detection. Further analysis shows that the correlation between MFS-TL characteristic and network state is non-deterministic in a datasets. 

According to observing the distributions of network characteristics, we found the normal state's values of most of network characteristics follow the Gaussian distribution. Thus, we define a simple rule for the statistical characteristic: the sampling MFS-TL is anomaly if the network characteristic value $c_i$ is lower than the $\psi$ in which the $\psi$ indicates the abnormal threshold for a given network characteristic sequence. So the threshold $\psi$ is related to the performance of the anomaly detection. Given a confidence interval $\theta$, the threshold $\psi$ can be computed by $\psi = \mu + \lambda \sigma$, where $\mu$ and $\sigma$ denote the mean and the standard deviation of one characteristic sequence $C=\{c_1, c_2,\dots,c_n\}$. And the $\lambda$ is the quantile of the normal distribution corresponding to the given confidence interval $\theta$\cite{meeker2014statistical}. In this paper, the $\varepsilon$ is 0.1 that confidence level of a MFS-TL characteristic sequence is $1-\varepsilon = 0.9$. Accordingly, the best detection metrics of CTU-9 MFS-TLs is the SPL, 0.9383, and the worst one is the diameter(max), 0.2469.

\subsection{Single characteristic-based anomaly detection} 
\label{subsec:single_characteristic_detection}

To analyze the evolution of MFS-TLs characteristics in all datasets, we found that there are the positive correlation, negative correlation and non-correlation between network characteristic and network states. For this reason, we use the intuitionistic fuzzy set (IFS) to quantify the non-deterministic correlations, that is let membership degree, non-membership degree, and hesitation degree of the IFS describe the positive correlation, negative correlation and non-correlation.

First, we calculate the anomaly detection accuracy ($Acc$) based on the IFS of single MFS-TL characteristic, as shown in Table \ref{tab:performance-1}. It can be found that the different characteristics have different detection performance in a dataset. For instance the $Acc$ values in ISOT-06, the best is 0.8520 from the node number and the worst is 0.2996 from the power-law. The accuracy of a MFS-TL characteristic is different with different datasets yet. For example, the MDR gets the best performance in ISOT-07, CICIDS-SSH, and CICIDS-DoS, but is the third worst metric in CTU-4. Thus, the results indicate that MFS-TL characteristics are inconsistent for anomaly detection performance. In Table \ref{tab:performance-1}, the best $Acc$ is 0.8520 for ISOT-06, 0.7148 for ISOT-07, 0.5755 for CTU-4, 0.8875 for CTU-9, 0.9019 for CICIDS-SSH and 0.8510 for CICIDS-DoS. The higher $Acc$ suggests that the correlation is more significant between network characteristic and network states. 
\begin{table*}[ht]
	\begin{center}
		\caption{The detection accuracy ($Acc$) based on single MFS-TL characteristic over different datasets. The bolded value is the best detection value. }\label{tab:performance-1}
		\begin{tabular}{ | l | c | c | c | c | c | c |}
			\hline
			Characteristic & ISOT-06         & ISOT-07         & CTU-4           & CTU-9           & CICIDS-SSH      & CICIDS-DoS      \\
			\hline
			Node number    & \textbf{0.8520} & 0.5064          & 0.473           & 0.8026          & 0.523           & 0.5104          \\
			Edge number    & 0.6968          & 0.4477          & 0.5423          & 0.8637          & 0.4062          & 0.4616          \\
			Mean Degree    & 0.7589          & 0.3813          & 0.4744          & \textbf{0.8875} & 0.4834          & 0.4454          \\
			Max Degree     & 0.6972          & 0.5613          & 0.4077          & 0.7107          & 0.8242          & 0.6177          \\
			MDR            & 0.5776          & \textbf{0.7148} & 0.3941          & 0.5299          & \textbf{0.9019} & \textbf{0.8510} \\
			K-core         & 0.7475          & 0.4363          & 0.449           & 0.887           & 0.5055          & 0.4973          \\
			Clique         & 0.7842          & 0.4162          & 0.3803          & 0.8383          & 0.4707          & 0.5047          \\
			Clustering     & 0.794           & 0.6879          & 0.5327          & 0.6508          & 0.6202          & 0.4831          \\
			Assortative    & 0.4708          & 0.5468          & 0.4818          & 0.8252          & 0.8384          & 0.8142          \\
			Entropy        & 0.7546          & 0.5197          & 0.5095          & 0.8583          & 0.4846          & 0.4972          \\
			SPL            & 0.5433          & 0.4538          & \textbf{0.5755} & 0.8668          & 0.8282          & 0.725           \\
			Diameter(max)  & 0.4255          & 0.4012          & 0.4805          & 0.7455          & 0.6792          & 0.5209          \\
			Diameter(mean) & 0.4237          & 0.4344          & 0.3858          & 0.7123          & 0.6944          & 0.5162          \\
			Power-law      & 0.2996          & 0.2593          & 0.4099          & 0.549           & 0.5398          & 0.5194          \\
			\hline
		\end{tabular}
	\end{center}
\end{table*}

\begin{figure}[h]
	\begin{center}
		\subfloat[ISOT-06 Node\label{sfig:ifs-06}]{%
			\includegraphics[width=0.495\columnwidth]{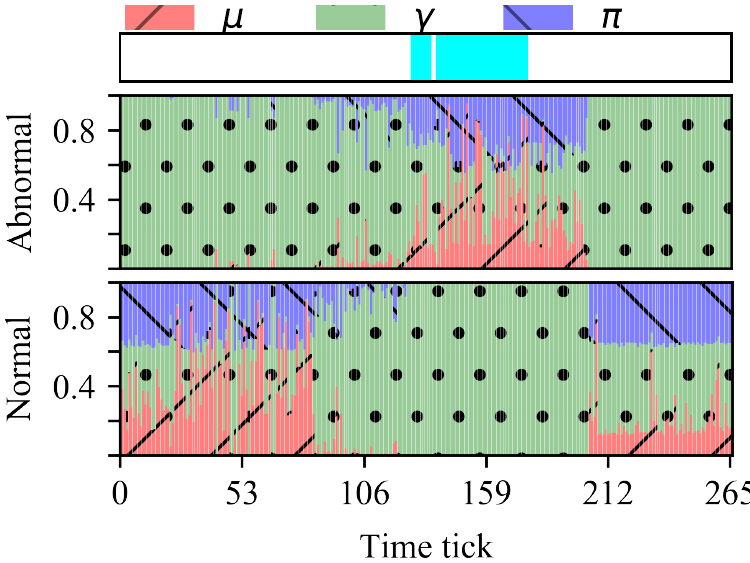}%
		}\hfill
		\subfloat[ISOT-07 MDR\label{sfig:ifs-07}]{%
			\includegraphics[width=0.495\columnwidth]{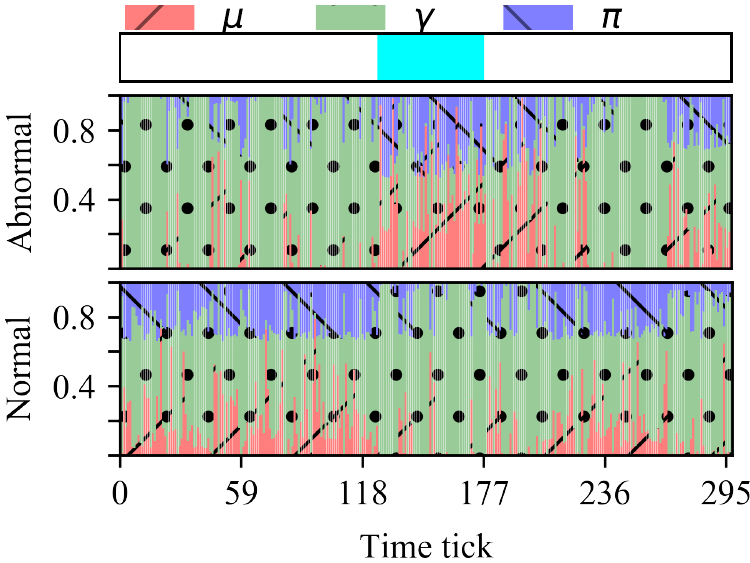}%
		}\\[-1em]
		\subfloat[CTU-4 SPL\label{sfig:ifs-4}]{%
			\includegraphics[width=0.495\columnwidth]{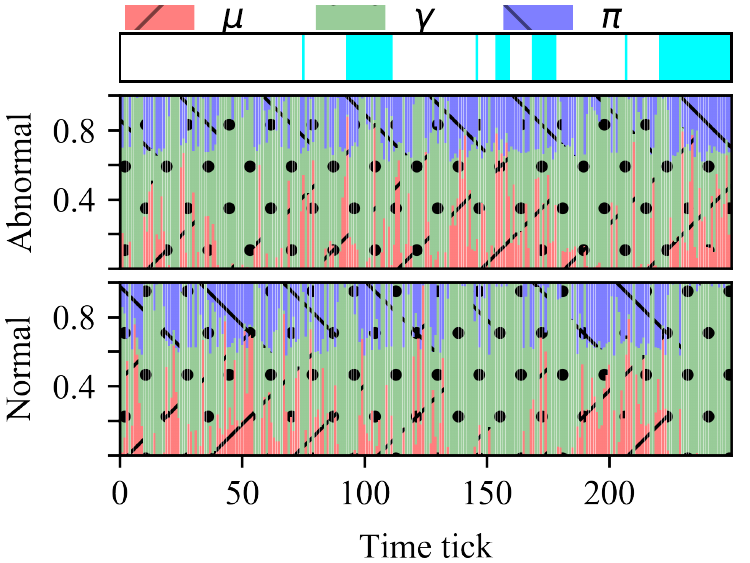}%
		}\hfill
		\subfloat[CTU-9 Mean degree\label{sfig:ifs-9}]{%
			\includegraphics[width=0.495\columnwidth]{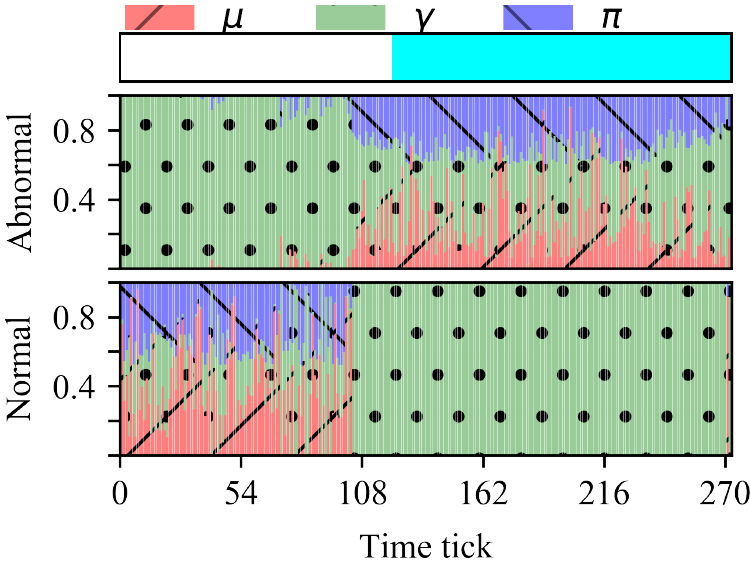}%
		}\\[-1em]
		\subfloat[CICIDS-SSH MDR\label{sfig:ifs-SSH}]{%
			\includegraphics[width=0.495\columnwidth]{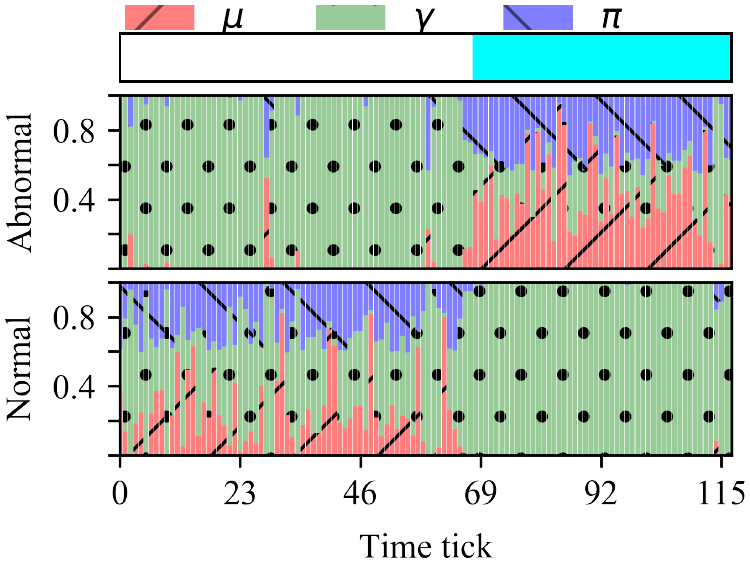}%
		}\hfill
		\subfloat[CICIDS-DoS MDR\label{sfig:ifs-DoS}]{%
			\includegraphics[width=0.495\columnwidth]{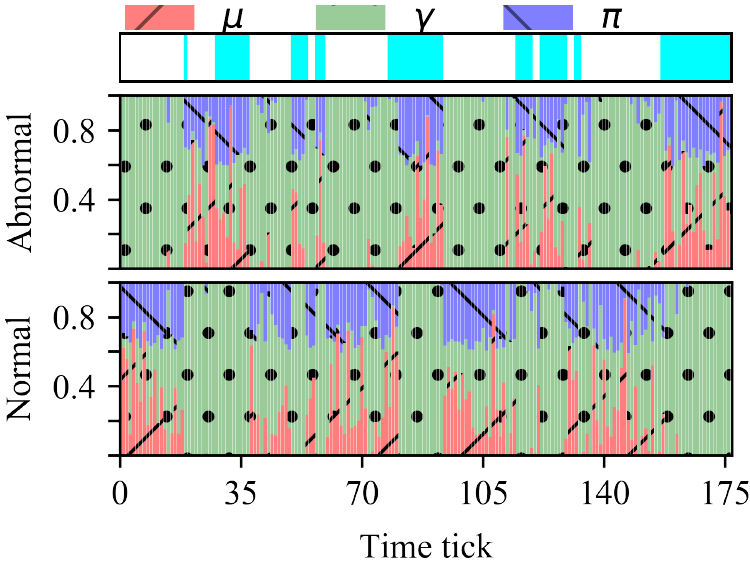}%
		}
		\caption{The distribution of membership degree $\mu$ (red bar with slash), non-membership degree $\gamma$ (green bar with black point) and hesitation degree $\pi$ (blue bar with backslash) over two linguistic variables as a function of time ticks for the best MFS-TL characteristic in each datasets. For each of the sequences of temporal network (a)-(f), it is abnormal, if the top ribbon is colored cyan at the time tick $i$, and vice versa. And the two bottom figures depict the abnormal and normal states, respectively.}
		\label{fig:ifs}
	\end{center}
\end{figure}

In order to exhibit the detection results of the best MFS-TL characteristic of each of datasets, we have plotted the IFS distributions of the best characteristic over two linguistic variables: \textit{abnormal} and \textit{normal}. The Fig.\ref{fig:ifs} describes the membership degree $\mu$, non-membership degree $\gamma$ and hesitation degree $\pi$ of sampling MFS-TLs. Apparently, the larger the red bar $\mu$ is, the higher probability the current network state should belong to this linguistic variable. For instance the IFS of the CTU-9 mean degree in Fig.\ref{sfig:ifs-9}, the anomaly happens continuously in the second half of the captured time. At the abnormal time ticks, the top sub-figure \textit{Abnormal} shows that the $\mu$ is greater than the $\gamma$ and $\pi$. In the bottom sub-figure \textit{Normal}, the values of $\gamma$ are in the range of $[0.9, 1]$. It indicates that the state of this network should belong to the linguistic variable \textit{abnormal}. Through comparative analysis with the anomaly event, we found that the detection results are accurate. Moreover, based on above analysis process, it can be seen that the others have similarity detection results.

In the above, using MFS-TL characteristic to describe network states is verified as shown in Table \ref{tab:performance-1} and Fig.\ref{fig:ifs}. The single MFS-TL characteristic IFS-based anomaly detection has a good detection performance for all datasets.

\subsection{Multiple characteristics-based ensemble detection} 
\label{subsec:ensemble_detection}

\begin{table*}[ht]
	\begin{center}
		\caption{ The performance of anomaly detection based on our method.}\label{tab:performance-2}
		\begin{tabular}{| p{0.20\columnwidth} | p{0.6\columnwidth} | c | c | c | c | c | c | c | c | }
			\hline
			Dataset    & Characteristic                                                                                     & TP & TF & FP & FN & $Acc$  & $Pre$  & $Rec$  & $F_1$  \\
			\hline
			ISOT-06    & Node, Edge, Mean degree, Clique, Clustering, Entropy, K-Core, Max degree.                          & 8  & 43 & 3  & 0  & 0.9444 & 0.7273 & 1.0000 & 0.8421 \\[4ex]
			ISOT-07    & Node, MDR, Clustering.                                                                             & 7  & 42 & 1  & 1  & 0.9608 & 0.8750 & 0.8750 & 0.8750 \\[4ex]
			CTU-4      & Entropy, Edge, Mean Degree.                                                                        & 18 & 22 & 10 & 1  & 0.7843 & 0.6423 & 0.9474 & 0.7660 \\[4ex]
			CTU-9      & SPL, Mean Degree, K-Core, Edge, Entropy, Assortative, Clique, Node, Diameter(mean), Diameter(max). & 34 & 26 & 1  & 0  & 0.9836 & 0.9714 & 1.0000 & 0.9855 \\[4ex] 
			CICIDS-SSH & SPL, MDR, Max degree, Assortative.                                                                 & 13 & 18 & 0  & 1  & 0.9688 & 1.0000 & 0.9286 & 0.9630 \\[4ex] 
			CICIDS-DOS & MDR, Assortative.                                                                                  & 15 & 21 & 1  & 0  & 0.9730 & 0.9375 & 1.0000 & 0.9677 \\
			\hline
		\end{tabular}
	\end{center}
\end{table*}
The above inconsistent performance of MFS-TL characteristic to network state motivates us to develope an ensemble method (IFSE-AD) for multiple MFS-TL characteristics to eliminate the impacts of the inconsistent and improve detect accuracy. In this paper, the performance of our method is analyzed by using the detection accuracy $Acc$, detection precision $Pre$, detection recall $Rec$, and $F_1$. Table \ref{tab:performance-2} shows the results of anomaly detection based on IFSE-AD. The $TP$, $TF$, $FP$, and $FN$ denote the number of corresponding detected instances in testing set. For the detection accuracy $Acc$, it can be seen that all are greater than 0.94 except for the CTU-4. That is because the distinction index $\tau$ of all MFS-TL characteristics in CTU-4 are lower than 0.5. In other words, there are weak correlations between CTU-4 characteristics and network states. The \textit{characteristics} field in Table \ref{tab:performance-2} shows the used characteristics in IFSE-AD method with the distinction index threshold $\tau_c = 0.5$. Noted that the used characteristics in CTU-4 are with the three best distinction index $\tau$, even though they are not greater than 0.5. Comparative analysis between $Pre$ and $Rec$, it is inferred that the big number of false positive $FP$ in detection results declines the detection precision $Pre$. It inspires us the future work about anomaly detection based on MFS-TL characteristic metrics. Fig.\ref{fig:ifs} shows that CTU-9, CICIDS-SSH, and CICIDS-DoS data have more abnormal instances, and in Table\ref{tab:performance-2} they have also higher $F_1$. It suggests that the more abnormal data improves the detection performance of IFSE-AD. Additionally, the study of the relationship between detection performance and the distinction index threshold $\tau_c$ denotes that the value of the $\tau_c$ have negligible influence to the detection performance.   

\begin{figure}[h]
	\begin{center}
		\includegraphics[width=1\columnwidth]{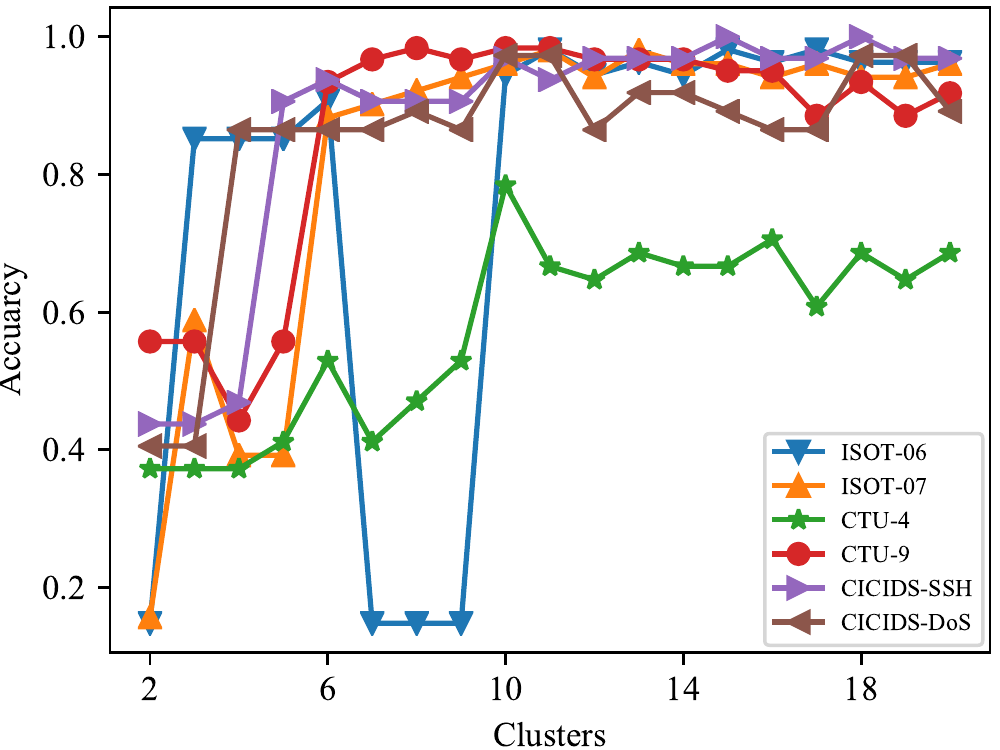}%
		\caption{The relationship between the accuracy $Acc$ and the clustering interval size.}
		\label{fig:a-cluster}
	\end{center}
\end{figure}

Besides, the relationship between the clustering interval size $c$ and the detection accuracy $Acc$ was studied. In Fig.\ref{fig:a-cluster}, it shows the trends of ISOT-07, CTU-04, CTU-09, CICIDS-SSH, and CICIDS-DoS follow logarithmic distribution. Except that the ISOT-06's accuracies have sudden changes at the clusters $c= 7, 8, 9$. It can be seen when the $c = 10$, there are good detection performances for all datasets overall. So in this paper the clustering interval size is set as 10 during partitioning the domain of discourse $D$.

\subsection{Evaluation} 
\label{subsec:evaluation}
\begin{table*}[h]
	\begin{center}
		\caption{Comparison of the accuracy/$F_1$ of anomaly detection by different methods.}\label{tab:result-compare}
		\begin{tabular}{ | l | c | c | c | c | c | c |}
			\hline
			Dataset    & K-Means+ID3     & Adaboost        & FIRE            & Gaussian-Dist   & IFS-AD          & IFSE-AD         \\
			\hline
			ISOT-06    & 0.7871 / 0.8370 & 0.9730 / 0.9793 & 0.8410 / 0.9695 & 0.9705 / 0.9811 & 0.8520 / 0.7022 & 0.9444 / 0.8421 \\
			ISOT-07    & 0.6519 / 0.7448 & 0.8640 / 0.9218 & 0.9347 / 0.9650 & 0.7222 / 0.7826 & 0.7148 / 0.5180 & 0.9608 / 0.8750 \\
			CTU-4      & 0.9898 / 0.9011 & 0.7120 / 0.8296 & 0.8280 / 0.9043 & 0.5517 / 0.6422 & 0.5755 / 0.4427 & 0.7843 / 0.7660 \\
			CTU-9      & 0.7861 / 0.7987 & 0.5590 / 0.7153 & 0.7245 / 0.7700 & 0.9383 / 0.9667 & 0.8875 / 0.9059 & 0.9836 / 0.9855 \\
			CICIDS-SSH & 0.7734 / 0.8043 & 0.8330 / 0.9441 & 0.8185 / 0.8592 & 0.8272 / 0.8940 & 0.9019 / 0.8832 & 0.9688 / 0.9630 \\
			CICIDS-DoS & 0.8055 / 0.8586 & 0.8880 / 0.9280 & 0.5020 / 0.6586 & 0.9483 / 0.9697 & 0.8510 / 0.8192 & 0.9730 / 0.9677 \\
			\hline
		\end{tabular}
	\end{center}
\end{table*}
Finally, the comparison of anomaly detection performance has been implemented by different methods. In this paper, we select three existed algorithms including K-Means+ID3\cite{Gaddam2007}, Adaboost\cite{Hu2008}, and FIRE\cite{Dickerson2000} in which the K-Means+ID3 and Adaboost, two of the most commonly used machine learning methods, detect the anomalies on the flow features extracted from the network traces, and the FIRE used the statistical metrics (e.g. the number of a port or the \textit{sdp}) of the network traces to construct the fuzzy logic to assess whether malicious activity is taking place. However, the FIRE and the methods based on MFS-TL is constructed on the sampled network traces with 1 minute sampling interval. So, the sizes of training set and testing set are far less than those of the K-Means+ID3 and Adaboost. Moreover, the Gaussian-Dist represents the anomaly detection method based on Gaussian distribution mentioned in the Section \ref{subsec:statistical-characteristics}. The IFS-AD represents the single characteristic-based anomaly detection method proposed in the Section \ref{subsec:single_characteristic_detection}. In this paper, the IFS-AD and IFSE-AD methods makes a fuzzification for every characteristic value to two states' linguist variables, and find which represents a best membership degree with multiple characteristics. In Table \ref{tab:result-compare}, the values denote the anomaly detection accuracy $Acc$ and $F_1$ separately. According to the comparison results, we found that the detection accuracy $Acc$ of the IFSE-AD are far better than that of K-Means+ID3, Adaboost, and FIRE. Although, the IFS-AD $Acc$ and $F_1$ are lower than the Gaussian-Dist, the IFS provide a good enough expression mechanism for the correlations between MFS-TL characteristic and network states that the IFSE-AD based on multiple characteristics has a better anomaly performance than the Gaussian-Dist. Additionally, we found that the methods based on the flow statistics and flow interaction have different sensitivity to the traffic data. Specifically, the K-Means+ID3, Adaboost, and FIRE based on the flow statistics have a better performance for what contains fewer anomaly activities. The Gaussian-Dist, IFS-AD, and IFSE-AD based on the flow interaction have a better performance for what contains more anomaly activities.

\section{Conclusion} 
\label{sec:conclusion}

In this work we have proposed the intuitionistic fuzzy set ensemble method (IFSE-AD) for anomaly detection of network traffic from the perspective of flow interaction. On the one hand, the multivariate flow similarity complex network model (MFS-TL) not only describe the interaction behaviors of large scale network flows, but also can monitor the dynamics of network traffic flows. On the other hand, our quantitative evaluation for network anomaly behaviors on publicly available network traffic datasets with ground truth show that building the IFSE-AD is effective in boosting detection performance. 

Initially, based on complex network theory, a complex network model, i.e. MFS-TL, is constructed by computing the multivariate flow similarity on temporal locality. Analyzing the relationships between MFS-TL characteristics, temporal locality window $\Delta w$, and multivariate flow similarity critical threshold $r_c$, an approach for parameter determination is established, i.e. finding the mutation point of MFS-TL entropy and SPL, and considering the communication latency. Thus, as shown in Fig.\ref{fig:params-ctu-9} the parameters are set as $\Delta w = 0.1s$ and $r_c = 0.65$ which exhibit the complexity and dynamic of network traffic. Observing the evolution of statistical characteristics of the MFS-TLs, three non-deterministic correlation types between MFS-TL characteristic and network state are defined, i.e. positive correlation (e.g. Fig.\ref{sfig:ch-ctu-9-edge}), negative correlation (e.g. Fig.\ref{sfig:ch-ctu-9-spl} and Fig.\ref{sfig:ch-ctu-9-assortative}), and non-correlation (e.g. Fig.\ref{sfig:ch-ctu-9-degree-max}).

Then, we introduced the intuitionistic fuzzy set (IFS) to quantify the correlation between MFS-TL characteristic and network state, i.e. membership degree of the IFS for positive correlation, non-membership degree of the IFS for negative correlation, and hesitation degree of the IFS for non-correlation. Furthermore, for a MFS-TL characteristic sequence, a IFS-based anomaly detection method (IFS-AD) is put forward to detect traffic anomalies. In IFS-AD, a Gaussian distribution-based membership function with a variable hesitation degree is designed to express the objectivity of intuitionistic fuzzification. The proposed distinction index resolves the mapping problem from multiple clustering intervals of the IFS to two states' linguistic variables. According to the accuracies in Table \ref{tab:performance-1}, we find the MDR obtains the best performance in ISOT-07 ($Acc$=0.7148), CICIDS-SSH ($Acc$=0.9019), and CICIDS-DoS ($Acc$=0.8510). Overall, every MFS-TL characteristic has different performance in a dataset, and the performance of a characteristic is also different with different datasets. It shows the inconsistent behaviors about MFS-TL characteristic to network state. Therefore, the intuitionistic fuzzy set ensemble method (IFSE-AD) is proposed to fuse the IFSs of multiple MFS-TL characteristics to eliminate the impacts of the inconsistent performance. The score function and precision function are used to sort the fused IFS. 

Finally we carried out extensive experiments on several network traffic datasets for anomaly detection. The detect accuracy/$F_1$ of the IFS-AD are 0.9444/0.8421 for ISOT-06, 0.9608/0.8750 for ISOT-07, 0.7843/0.7660 for CTU-4, 0.9836/0.9855 for CTU-9, 0.9688/0.9630 for CICIDS-SSH, and 0.9730/0.9677 for CICIDS-DoS. The results demonstrate the superiority of our method to state-of-the-art approaches, validating the effectiveness of our method. Additionally, the methods based on the flow statistics and flow interaction have different sensitivity to the data: the K-Means+ID3, Adaboost, and FIRE based on the flow statistics have a better performance in network traffic containing fewer anomaly activities. The Gaussian-Dist, IFS-AD, and IFSE-AD based on the flow interaction have a better performance in which contains more anomaly activities. All source code of our methods, the data used in this work, and the more charts about parameters analysis and statistical characteristics of MFS-TL are shared openly at \underline{\url{http://file.mervin.me/project/internet-mfstl-ad}}

\bibliographystyle{IEEEtran}
\bibliography{refs}

\EOD
\end{document}